%% file: main.tex
\documentclass[a4paper]{article}

\usepackage{graphicx}      
\usepackage[backend=biber, language=auto, maxnames=3]{biblatex}
\usepackage{amsmath,amssymb,amsthm}
\usepackage{xcolor}
\usepackage{microtype}
\usepackage{fancyhdr}

\input{macros.tex}

\newcommand{\citep}{\cite}

\fancypagestyle{FirstPage}{
\fancyhf{}
\lfoot{\small \copyright2023.
This work has been accepted to IFAC for publication under a Creative Commons Licence CC-BY-NC-ND.}
}

\newtheorem{thm}{Theorem}
\newtheorem{cor}[thm]{Corollary}
\newtheorem{fact}[thm]{Fact}

\author{Lev-Arcady Sellem%
\thanks{%
Laboratoire  de Physique de l’Ecole Normale Sup\'{e}rieure, %
	Mines Paris-PSL, %
	CNRS, %
	ENS-PSL, %
	Inria, %
	Sorbonne Universit\'{e}, %
	Université PSL, %
	Paris, France.%
} \,%
\thanks{Corresponding author: lev-arcady.sellem@inria.fr}
\and Rémi Robin$^*$ \and Philippe Campagne-Ibarcq$^*$ \and Pierre Rouchon$^*$}

\title{%
	Stability and decoherence rates %
	of a GKP qubit protected by dissipation%
	}
\begin{document}

\maketitle
\thispagestyle{FirstPage}

\begin{abstract}
	We analyze an experimentally accessible Lindblad master equation for a
	quantum harmonic oscillator. It approximately stabilizes finite-energy periodic grid
	states called Gottesman-Kitaev-Preskill (GKP) states, that can be
	used to encode and protect a logical qubit. We give
	explicit upper bounds for the  energy  of the solutions of
	the Lindblad master equation. Using three periodic observables to define
	the Bloch sphere coordinates of a  logical qubit, we show
	that their dynamics is governed  by a diffusion  partial differential
	equation on a 2D-torus with a Witten Laplacian.	We
	show that the evolution of these logical coordinates is exponentially
	slow even in presence of small diffusive noise processes along
	the two quadratures of the phase space.	Numerical simulations indicate
	similar results	for other physically relevant noise processes.
\end{abstract}


\section*{Introduction}
\input{intro.tex}

\section{Lindblad master equation}
\label{sec:lindblad_equation}
\input{section_lindblad_equation.tex}

\section{Stability analysis}
\label{sec:stability}
\input{section_energy.tex}
\section{Explicit decoherence rates}
\label{sec:explicit_rates}
\input{section_decoherence_rates.tex}

\clearpage

\section{Conclusion}
\label{sec:conclusion}
\input{conclusion.tex}

\paragraph{Acknowledgments.}
	We thank Alain Sarlette for useful discussions and comments.
This project has received funding from the European Research Council (ERC)
	under the European Union’s Horizon 2020 research and innovation program
	(grant agreement No. 884762).
\printbibliography

\clearpage

\appendix
\section{Proof of Theorem $1$}
\label{adx:energy_proof}
\input{appendix_energy.tex}

\clearpage
\section{Proof of Theorem $3$}
\label{adx:eigs_proof}
\input{appendix_eigs.tex}

\end{document}

%% file: macros.tex
\usepackage{upgreek}
\usepackage{svg}

\let\GGamma\Gamma
\renewcommand{\Gamma}{}

\DeclareMathOperator{\trace}{Tr}
\DeclareMathOperator{\sign}{Sign}
\DeclareMathOperator{\hc}{h.c.}
\newcommand{\qop}{{\bf q}}

\newcommand{\hop}{{\bf h}}

\newcommand{\Pop}{{\bf p}}

\newcommand{\X}{{\bf X}}
\newcommand{\Z}{{\bf Z}}
\newcommand{\Y}{{\bf Y}}

\newcommand{\bsigma}{\boldsymbol{\sigma}}

\newcommand{\VV}{{\bf V}}
\newcommand{\EE}{{\bf E}}

\newcommand{\II}{{\bf 1}}

\newcommand{\N}{{\bf N}}

\newcommand{\LL}{{\bf L}}

\usepackage{hyperref}

\newcommand{\cA}{\mathcal A}
\newcommand{\cD}{\mathcal D}
\newcommand{\oa}{{\bf a}}
\renewcommand{\dag}{\dagger}
\renewcommand{\hc}{\textrm{h.c.}}
\newcommand{\ie}{\emph{i.e.}}
\newcommand{\eg}{\emph{e.g.}}

%% file: intro.tex
Quantum error correction (QEC) is the main
missing block in the construction of a universal quantum computer
able to perform useful algorithms.
Indeed, all currently available quantum computing platforms
suffer from high decoherence (\ie{} uncontrolled noise) limiting the size
and duration of computations that can be run before information is lost.
\\

Seen from above,
quantum error correcting codes,
such as
the color code~\cite{bombinTopologicalQuantum2006}
or surface code~\cite{bravyiQuantumCodes1998}
(see also~\cite{qeczoo} for a more exhaustive overview),
encode a logical qubit in a collection of physical qubits,
thus embedding it in a Hilbert space of larger dimension
in a way such that
typical physical errors translate to shifts between two-dimensional subspaces
that can be later corrected \citep{nielsenchuang}.
However, these codes typically work provided that
the noise affecting physical qubits can be kept below a given threshold
(whose exact value depends on the specific code under consideration);
despite considerable progress
in reducing noise levels and increasing code thresholds,
this condition is yet to bet observed in experiments
\citep{chenExponentialSuppression2021a,krinnerRealizingRepeated2022}.

Bosonic encodings provide a promising alternative approach:
instead of using the Hilbert space of a collection of noisy physical qubits,
a logical qubit is encoded in the infinite-dimensional state space of a quantum harmonic oscillator.
The goal can either be to fully replace QEC codes,
or more realistically to reach error levels below the threshold of a given QEC code.
In the latter case, full quantum error correction can then be achieved
through a form of code concatenation:
the innermost code is actually a bosonic encoding scheme,
that can be seen as a hardware-efficient error correcting code,
while the (possibly multiple) next codes are traditional QEC codes
(see \eg{}~%
\cite{guillaudRepetitionCat2019, chamberlandBuildingFaultTolerant2022, nohLowOverheadFaultTolerant2022}).
In a bosonic encoding,
the two-dimensional subspace used to encode the qubit must be chosen so that
typical noise processes (depending on the specific physical implementation)
minimally disturb the encoded information.
This usually involves exploiting exotic states of the oscillator;
for instance, Fock states for
the binomial code \citep{michaelNewClass2016},
superposition of coherent states for
the cat code \citep{cochraneMacroscopicallyDistinct1999}
or periodic grid states for
the Gottesman-Kitaev-Preskill (GKP) code
\citep{gottesmanEncodingQubit2001}.
Experimental stabilization and manipulation of these exotic states can present
a significant experimental challenge.
\\

In the case of the GKP code,
stabilization of the involved grid states
has been experimentally demonstrated in
superconducting circuits~\cite{campagne-ibarcqQuantumError2020a}
and trapped ions~\cite{deneeveErrorCorrection2020};
these experiments rely on repeated interactions with an ancillary system
to perform a discrete-time feedback control loop.
This strategy is limited by the strength of the coupling to the ancilla,
the propagation of ancilla errors to the qubit
and the speed of the discrete-time feedback loop.
Alternately, autonomous and continuous-time stabilization could be achieved
with a carefully engineered coupling to a dissipative environment
such that the system naturally relaxes to the exotic states encoding logical information.
This has already been experimentally demonstrated for the cat code%
~\cite{lescanneExponentialSuppression2020},
but not for the GKP code.

In~\cite{royerStabilizationFiniteEnergy2020}, the authors
numerically studied 
the potential of a Lindblad master equation with two dissipators
for the autonomous stabilization of finite-energy GKP states,
but did not investigate the stability or
convergence rates of the proposed dynamics.
Additionally, it is unclear whether the dissipators they introduce
can be experimentally realized with current technologies;
they proposed instead to approximate them by a non-autonomous scheme
based on a coupling with an ancilla qubit and repeated measurements of the ancilla followed
by feedback pulses.
 
In~\cite{sellemExponentialConvergence2022}, the authors
provided explicit convergence rates based on a Lyapunov approach
for a Lindblad master equation with four dissipators
stabilizing finite-energy GKP states,
and provided numerical evidence of the potential of their dynamics
for the protection of quantum information.
To the best of our knowledge,
no similar results on stability and convergence towards GKP states are available
for a Lindblad master equation with only two dissipators.
The dissipators considered also seem
out of reach for current technologies;
however, 
an implementation of approximate versions of these dissipators
in a circuit-QED device under realistic physical constraints
was later proposed in~%
\cite{sellemGkpImplementation},
along with numerical evidence that these approximate dissipators
still feature promising properties for the encoding, protection and operation of
quantum bits (qubits).
Here, we analyze the mathematical properties of the dynamics induced by these four
approximate dissipators.
\\

In Sec.~\ref{sec:lindblad_equation}
we define the Lindblad master equation under study.
In Sec.~\ref{sec:stability}
we prove stability of the Lindblad dynamics
and derive explicit bounds on the energy of the system.
In Sec.~\ref{sec:explicit_rates}
we show that, when considering quadrature noise only,
decoherence rates of the encoded quantum information
can be computed explicitly without actually solving the Lindblad equation.
For more realistic noise models, we provide numerical simulations
indicating robustness of the observed protection.
Concluding remarks and possible further developments are gathered in Sec.~\ref{sec:conclusion}.
Details of the computations can be found in appendix.

%% file: section_lindblad_equation.tex
In~\cite{sellemGkpImplementation}
two different Lindblad master equations,
on the Hilbert space $\mathcal H = L^2(\mathbb R,\mathbb C)$ of a quantum harmonic oscillator,
are considered
for the 
stabilization of finite-energy GKP grid states of the oscillator.
These two Lindblad equations
involve four (respectively six) dissipators
corresponding to square (respectively hexagonal) grid states.
We can re-derive both families of dissipators
from the finite-energy dissipators
introduced in~\cite{sellemExponentialConvergence2022}.
Take $\eta>0$, $\Delta>0$, $M\in\mathbb N^*$;
consider the Hermitian phase-space operators of a quantum harmonic oscillator
$\qop$
and $\Pop$ satisfying the commutation relation $[\qop, \Pop] = i$
and the photon-number operator
$\N = \frac12 (\qop^2+\Pop^2 - \II)$.
Following~\cite{sellemExponentialConvergence2022},
we can define the operator
\begin{equation}
\VV_0 = \EE_\Delta \left( e^{i\eta\qop} - \II \right) \EE_\Delta^{-1}
\end{equation}
with $\EE_\Delta = e^{-\Delta \N}$ a regularizing operator.
This operator, as well as its rotations of angle
$\theta_k = k\pi/M$ in phase-space defined by
\begin{equation}
	\VV_k = e^{i\theta_k \N} \, \VV_0 \, e^{-i\theta_k \N}, \quad 0\leq k\leq 2M-1,
\end{equation}
were built to identically vanish on finite-energy GKP grid-states.

Now, using the Baker-Campbell-Hausdorff formula we get
\begin{equation}
\EE_\Delta \, \qop \, \EE_\Delta^{-1} = \cosh(\Delta) \qop + i\sinh(\Delta)\Pop
\end{equation}
giving the formal approximation of $\VV_0$ to first order in $\Delta$:
\begin{align}
\VV_0 &= e^{-\frac12\eta^2\cosh(\Delta)\sinh(\Delta)} \,
	e^{i\eta\cosh(\Delta)\qop} \,
	e^{-\eta\sinh(\Delta)\Pop} - \II\\
	&\simeq \cA \, e^{i\eta\qop} \left( \II-\epsilon\Pop\right) - \II
	\triangleq \LL_0
	\label{eq:def_l0}
\end{align}
with $\epsilon = \eta\sinh(\Delta)$ and $\cA = e^{-\epsilon\eta/2}$.
For a quantum harmonic oscillator described by its density operator $\rho$
(\ie{} a trace-class hermitian positive operator on $L^2(\mathbb R,\mathbb C)$
with $\trace(\rho) = 1$)
we consider the Lindblad master equation
\begin{equation}
\frac d{dt} \rho
	= \sum_{k=0}^{2M-1} \cD[\LL_k] \rho
	\label{eq:lindblad_generic}
\end{equation}
where we defined a family of Lindblad operators
\begin{equation}
	\LL_k = e^{i\theta_k \N} \, \LL_0 \, e^{-i\theta_k \N}
\end{equation}
corresponding to rotations of $\LL_0$
of angle $\theta_k$
in phase-space,
and the dissipator $\cD[\LL]$ is defined as
\begin{equation}
	\cD[\LL](\rho)
	= \LL \rho  \LL^\dag - \frac12 \left( \LL^\dag\LL \, \rho + \rho\, \LL^\dag \LL\right).
\end{equation}
Within this general framework,
we recover the Lindblad master equation stabilizing a square GKP codespace
(\ie{} the space spanned by GKP states corresponding to square grid states)
with the choice
\begin{equation}
	\label{eq:m_eta_square}
	M = 2, \quad \eta = 2\sqrt\pi
\end{equation}
while
we recover the Lindblad master equation stabilizing a hexagonal GKP codespace
with the choice
\begin{equation}
	\label{eq:m_eta_hexa}
	M = 3, \quad \eta = 2\sqrt{\frac{2\pi}{\sqrt3}}.
\end{equation}
Note that our framework could be easily adapted
to encompass variants of the square GKP code,
based on rectangular or parallelogrammatic grids,
as well as to encode a qudit of dimension $d$
instead of a qubit (corresponding to $d=2$).
For instance, encoding a qudit in a square grid
corresponds to picking $M=2, \, \eta = \sqrt{2d\pi}$.
We expect the results exposed in this paper to be easily adapted
to such generalized geometries
but leave the study of this question to future work.
\\

At this stage, the main impact of the approximation made in Eq.~\eqref{eq:def_l0}
is that,
contrary to the Lindblad operators considered in~\cite{sellemExponentialConvergence2022},
the Lindblad operators $\LL_k$ do not commute
and fail to identically vanish on finite-energy GKP grid-states;
as a result, we can no longer hope to stabilize these states exactly.
In other words, the approximation made to obtain an experimentally accessible
dynamics introduces intrinsic residual decoherence.
However, we show in the next sections that the energy of the solutions
to Eq.~\eqref{eq:lindblad_generic} remain bounded along trajectories
and that
we can compute decoherence rates of a logical qubit encoded in our system
in presence of additionnal noise channels introduced in the model;
in particular, setting the noise strength to zero,
we will see that
the intrinsic decoherence rate due to the imperfect stabilization of grid-states
is exponentially small in $1/\epsilon$.

%% file: section_energy.tex
\label{sec:energy_estimates}

We can compute explicit bounds on the
energy,
defined as the expectation value \( \langle \N \rangle = \trace(\N\rho) \)
of the photon number operator $\N$,
along trajectories of a density operator
governed by Eq.~(\ref{eq:lindblad_generic}).
Here, we momentarily forget the goal of encoding logical information
and only focus on stability properties of the Lindblad dynamics.
In particular, we provide energy bounds for arbitrary values of $\eta$ and $M$,
while only the two choices in Eq.~(\ref{eq:m_eta_square})--(\ref{eq:m_eta_hexa}),
used to encode a qubit,
will be considered in the next section.

We obtain
\emph{a priori} estimates on the solution to
Eq.~(\ref{eq:lindblad_generic})
by
formal computations, led
as if the dimension of the underlying Hilbert space
were finite
(ignoring in particular any potential issue
related to the domains of the involved unbounded operators).
We plan to exploit these estimates as a first step towards
a fully rigorous mathematical analysis
in future publications.
\\

\begin{thm}
	Assume that $\trace(\N \rho_0) <+\infty$,
	$\epsilon\eta/2 < 0.4$
	and the evolution of $\rho_t$ is governed by the Lindblad equation~(\ref{eq:lindblad_generic}).
	Then, for any $r\in (0,1)$ and for all $t\geq 0$:
	\begin{equation}
		\label{eq__energyboundderiv}
		\frac d{dt}\trace(\N \rho_t) \leq
		-\lambda(r,\epsilon,\eta) \trace(\N\rho_t) + \mu(r,\epsilon,\eta)
	\end{equation}
	where
	\begin{align}
		\label{eq__lambda_coef}
		\lambda(r,\epsilon,\eta) &=
		2M \Gamma \cA \, r \epsilon \eta \, \phi_{\epsilon,\eta} > 0, \\
		\label{eq__mu_coef}
		\mu(r,\epsilon,\eta) &= %
		M\Gamma\cA \Big[
		(\epsilon^2+\eta^2)\cA + \frac{\epsilon\eta^3}{2}
			+ \frac{\epsilon}{2\eta(1-r)\, \phi_{\epsilon,\eta}}
		 - r \epsilon\eta \, \phi_{\epsilon,\eta} \Big],\\
		\phi_{\epsilon,\eta} &= (2-\epsilon\eta/2) \cA - 1 = (2-\epsilon\eta/2) \, %
		e^{-\epsilon\eta/2}-1.
	\end{align}

\label{thm:energy_estimates}
\end{thm}
The proof of Theorem~\ref{thm:energy_estimates}
is presented in Apprendix~\ref{adx:energy_proof}.
We can use the differential inequality of Theorem~\ref{thm:energy_estimates}
to obtain an estimate on the energy at all times through
Grönwall's lemma:
\begin{cor}
	Under the same hypothesis
	as Theorem~\ref{thm:energy_estimates},
	the following bound on the energy holds:
\begin{equation}
	\begin{split}
		\trace(\N\rho_t) \leq \, &
		e^{-\lambda(r,\epsilon,\eta) t} \trace(\N\rho_0)
		+ \left(1-e^{-\lambda(r,\epsilon,\eta) t}\right) C(r,\epsilon,\eta)
	\end{split}
	\label{eq__energyboundvalue}
\end{equation}
where
$C(r,\epsilon,\eta) = \mu(r,\epsilon,\eta)/\lambda(r,\epsilon,\eta)$.
\end{cor}

In the limit $\epsilon\rightarrow 0^+$, we get the following asymptotics:
\begin{align}
		\lambda(r,\epsilon, \eta) &\sim 2r \,  M \Gamma \, \epsilon \eta,\\
		C(r,\epsilon, \eta) &\sim \frac{1}{2r} \, \frac\eta\epsilon.
		\label{eq:estimate_energy_C}
\end{align}

Note that here, $r\in(0,1)$ is a free parameter which we need to introduce in the proof of
Theorem~\ref{thm:energy_estimates}.
When one is only interested in the asymptotic value of the bound on energy
for small $\epsilon$,
the value $r=1-\sqrt\epsilon$
leads to the simpler asymptotics
		$\lambda(\epsilon, \eta) \sim 2 M \Gamma \epsilon \eta$,
		$C(\epsilon, \eta) \sim \eta/(2\epsilon)$.
On the other hand, the hypothesis $\epsilon\eta/2 < 0.4$ is only a sufficient condition
to ensure $\phi_{\epsilon,\eta} = (2-\epsilon\eta/2)\cA-1 >0$ and can be safely mentally replaced by
\emph{for small enough~$\epsilon$}.
In particular, it is satisfied in every numerical simulation presented here.
\\

In addition to proving stability of the dynamics,
the above estimates are useful
from a numerical point of view.
The usual method for the simulation of a quantum harmonic oscillator
relies on a Galerkin approximation in the so-called Fock basis,
an orthonormal basis of $L^2(\mathbb R,\mathbb C)$
formed by eigenvectors of $\N$.
In our case,
the energy estimates of Eq.~(\ref{eq__energyboundvalue})
give a rationale for choosing an adapted truncation
in this basis.

%% file: section_decoherence_rates.tex
A (logical) qubit is a density operator $\rho_L$ on the Hilbert space $\mathbb C^2$,
usually represented by its coordinates $(x,y,z)$ on the Bloch sphere as
\begin{equation}
	\begin{aligned}
		\rho_L = &\frac{\II + x \, \bsigma_x + y \, \bsigma_y + z \, \bsigma_z}2, \quad x^2+y^2+z^2 \leq 1
	\end{aligned}
\end{equation}
where $\bsigma_x, \, \bsigma_y, \, \bsigma_z$ are the usual Pauli matrices
and
\[ x = \trace(\rho_L \, \bsigma_x),\,
y = \trace(\rho_L \, \bsigma_y),\,
z = \trace(\rho_L \, \bsigma_z). \]

To encode a qubit in the state of a quantum harmonic oscillator,
described by its density operator $\rho$ on the Hilbert space
$L^2(\mathbb R,\mathbb C)$,
we must specify an encoding map
\begin{equation}
	\mathcal E \colon \rho \mapsto \rho_L.
\end{equation}
When the harmonic oscillator evolves through a Lindblad equation
\begin{equation}
	\frac d{dt} \rho = \mathcal L(\rho)
\end{equation}
we want to compute the associated evolution of the encoded logical qubit,
for instance
\begin{equation}
	\dot x = \trace\left( \frac d{dt} \mathcal E(\rho) \, \bsigma_x\right).
\end{equation}

In our case, \cite{sellemGkpImplementation}
used an encoding map $\mathcal E$ adapted to the study of grid states
by introducing the following so-called \emph{generalized} Pauli operators on $L^2(\mathbb R,\mathbb C)$:
\begin{align}
	\Z &= \sign(\cos(\tfrac\eta2 \qop_1))\\
	\X &= \sign(\cos(\tfrac\eta2 \qop_2))\\
	\Y &= -i \, \Z \X
	\label{eq:gen_pauli}
\end{align}
where $\eta=2\sqrt\pi$, $\qop_1 = \qop$, $\qop_2=\Pop$ for a square GKP codespace
and $\eta=2\sqrt{\frac{2\pi}{\sqrt3}}$, $\qop_1 = \qop$, $\qop_2= -\frac12\qop + \frac{\sqrt3}2\Pop$
for a hexagonal GKP codespace.
In both cases, one can check that this defines
bounded hermitian operators with spectrum $\{-1,1\}$, anti-commuting with each other,
and satisfying the usual Pauli algebra
\begin{align}
	\X \Y = i \, \Z, \quad
	\Y \Z = i \, \X, \quad
	\Z \X = i \, \Y.
\end{align}
The encoding is then defined as
\begin{align}
	x = \trace(\X \rho), \quad
	y = \trace(\Y \rho), \quad
	z = \trace(\Z \rho).
\end{align}

Let us now assume that $\rho$ evolves according to Eq.(\ref{eq:lindblad_generic}),
possibly enriched with additional error terms modeling for instance experimental imperfections:
\begin{equation}
	\frac d{dt}\rho =
	\sum_{k=0}^{2M-1} \cD[\LL_k](\rho)
	+ \kappa \, \mathcal L_{\textrm{noise}}(\rho)
	\label{eq:lindblad_with_noise}
\end{equation}
In the noiseless case ($\mathcal L_{\textrm{noise}} = 0$)
or when considering quadrature noise
($\mathcal L_{\textrm{noise}} = \cD[\qop] + \cD[\Pop]$)
we show that the decay rate of the logical coordinates
$(x,y,z)$ along trajectories can be
computed without
solving the Lindblad equation~(\ref{eq:lindblad_with_noise}).
More precisely, we exploit the fact
that the generalized Pauli operators defined in Eq.(\ref{eq:gen_pauli})
are products of periodic functions of $\qop_1$ and $\qop_2$ to obtain the following
result:
\begin{thm}[Square case]
	Let $M=2$, $\eta=2\sqrt\pi$, and
	$f,g$ be two real-valued $2\pi$-periodic functions.
	Define the periodic function
		$h(\theta_1,\theta_2) = f(\theta_1)g(\theta_2)$
	and the periodic operator
	\begin{equation}
	\hop = h(\tfrac\eta2\qop, \tfrac\eta2\Pop) =
f(\tfrac\eta2\qop) \, g(\tfrac\eta2\Pop)
	\end{equation}
	and assume
	that $\rho$ evolves through Eq.(\ref{eq:lindblad_generic})
	augmented by quadrature noise:
	\begin{equation}
		\frac d{dt} \rho_t = \sum_{k=0}^{2M-1} \cD[\LL_k](\rho_t)
				+ \kappa \, \cD[\qop](\rho_t)
				+ \kappa \, \cD[\Pop](\rho_t).
	\end{equation}
	Then, the expectation value of $\hop$ satisfies:
	\begin{equation}
		\frac d{dt} \trace(\hop \rho_t) = -\cA \epsilon\eta
		\trace\left( \mathcal L_\sigma h (\tfrac\eta2\qop, \tfrac\eta2\Pop) \, \rho_t \right)
		\label{eq:evolution_expectation_eigs}
	\end{equation}
	with
	\begin{align}
		\cA &= e^{-\epsilon\eta/2}, \\
		\sigma &\equiv \sigma(\epsilon,\eta,\kappa)
		= \frac{\cA\epsilon\eta}4 + \frac{\kappa\eta}{8\cA\epsilon}, \\
		\mathcal L_\sigma &= \mathcal T_\sigma \otimes \II + \II \otimes \mathcal T_\sigma,
		\label{eq:l_from_t}\\
		(\mathcal T_\sigma f)(\theta) &= \sin(2\theta) f'(\theta) - \sigma f''(\theta).
	\end{align}
	\label{thm:eigs_square}
\end{thm}
The proof of Theorem~\ref{thm:eigs_square} is presented in Apprendix~\ref{adx:eigs_proof}.
As a consequence,
we can compute the evolution of $\trace(\hop \rho_t)$
if we know the eigenvalues and eigenfunctions of $\mathcal L_\sigma$,
which we can deduce from that of $\mathcal T_\sigma$.
One can check that $\mathcal T_\sigma$
is self-adjoint
on $H^2_{per}(-\pi,\pi)$
for the scalar product
\begin{equation}
	\langle f, g \rangle_\sigma =
	\int_{-\pi}^\pi e^{-\frac{1-\cos(2\theta)}{2\sigma}} f(\theta) g(\theta) d\theta
\end{equation}
and that
\begin{equation}
	\langle f, \mathcal T_\sigma f\rangle_\sigma = \sigma \, \langle f', f'\rangle_\sigma,
\end{equation}
from which we can deduce from classical arguments
(see \eg{} \cite{ZettlBook2005})
that the spectrum of
$\mathcal T_\sigma$ is discrete, real and non-negative and can thus be denoted as
\begin{equation}
	0 = \mu_{0,\sigma} \leq \mu_{1,\sigma} \leq \mu_{2,\sigma} \leq \ldots
\end{equation}
where $\mu_{0,\sigma} = 0$ is a trivial eigenvalue
associated to a constant eigenfunction.
The results of~\cite{michelSmallEigenvalues2019a}
on Witten Laplacians
can be applied to $\mathcal T_\sigma$
to obtain its next eigenvalues
in the regime $\sigma\rightarrow 0^+$:
\begin{fact}
	\begin{align}
		&\mu_{1,\sigma} = \left( \frac4\pi + o(1) \right) e^{-1/\sigma}, \\
		& \exists \upepsilon_0>0, \forall \sigma>0, \quad
			\mu_{2,\sigma} \geq \upepsilon_0.
		\label{eq:eigs_michel_square}
	\end{align}
\end{fact}
By combining Eq.(\ref{eq:eigs_michel_square}) and~(\ref{eq:l_from_t}),
we deduce the eigenvalues of $\mathcal L_\sigma$, that we can denote
\begin{equation}
	0 = \lambda_{0,\sigma} \leq \lambda_{1,\sigma} \leq \lambda_{2,\sigma} \leq \ldots
\end{equation}
\begin{align}
	\lambda_{1,\sigma} &= \mu_{0,\sigma} + \mu_{1,\sigma} =
		\left( \frac4\pi + o(1) \right) e^{-1/\sigma}, \label{eq:asymptotic_lambda1}\\
	\lambda_{2,\sigma} &= \mu_{1,\sigma} + \mu_{0,\sigma} =
		\left( \frac4\pi + o(1) \right) e^{-1/\sigma},\\
	\lambda_{3,\sigma} &= \mu_{1,\sigma} + \mu_{1,\sigma} =
		\left( \frac8\pi + o(1) \right) e^{-1/\sigma},\\
	\lambda_{4,\sigma} &= \mu_{2,\sigma} + \mu_{0,\sigma} \geq \upepsilon_0.
\end{align}
In particular, the contribution of eigenfunctions associated to eigenvalues
$\lambda_{k,\sigma}$ for $k\geq4$ decays
at a rate exponentially larger, as a function of $\sigma$,
than
that of the three eigenfunctions associated to the first three non-zero eigenvalues.
Going back to the result of Theorem~\ref{thm:eigs_square},
we see that after an initial transient on a typical timescale
$\tau_{trans} = (\cA \epsilon\eta)^{-1}$,
the expectation value of any periodic observable
(in particular those defining logical coordinates)
decay on an exponentially slower timescale
$\tau_{decay} = \left(\frac 4\pi \cA \epsilon \eta \, e^{-1/\sigma} \right)^{-1}$.
\\

Unfortunately, for noise models more generic than quadrature noise,
we could not adapt the above analysis,
as the proof of Theorem~\ref{thm:eigs_square}
is highly dependent on commutation relations between Lindblad operators entering the dynamics
and periodic operators.
However,
to assess the robustness of the proposed GKP encoding with respect to various
noise processes,
we compared on Figure~\ref{fig:numerics}
the actual decoherence rates
extracted from full Lindblad simulations of Eq.(\ref{eq:lindblad_with_noise})
for different choices of $\mathcal L_{\textrm{noise}}$,
with the asymptotic value of  $\lambda_{1,\sigma}$
obtained for quadrature noise.
While each type of noise yields quantitatively different decoherence rates,
we observe a qualitatively similar decay of the decoherence rate
as a function of the noise strength $\kappa$.
Note the nonzero value of the decoherence rate in absence of noise ($\kappa=0$),
attributed in Section~\ref{sec:lindblad_equation} to the fact
that the approximate finite-energy Lindblad operators $\LL_k$ do not commute:
the approximation made to obtain an experimentally accessible
dynamics leads to intrinsic residual decoherence,
at a rate $1/\tau_{decay}= \tfrac{4\cA\epsilon\eta}\pi \, e^{-4/(\cA\epsilon\eta)}$
exponentially small in $1/\epsilon$.
\\

Finally, a similar analysis can be led for the hexagonal GKP code:
\begin{thm}[Hexagonal case]
	Let $M=3$, $\eta=2\sqrt{\frac{2\pi}{\sqrt3}}$, and
	$f,g$ be two real-valued $2\pi$-periodic functions.
	Define the periodic function
		$h(\theta_1,\theta_2) = f(\theta_1)g(\theta_2)$
	and the periodic operator
	\begin{equation}
	\hop = h(\tfrac\eta2\qop, \tfrac\eta2\qop_2) =
f(\tfrac\eta2\qop) \, g(\tfrac\eta2\qop_2)
	\end{equation}
	with $\qop_2 = -\frac12 \qop + \frac{\sqrt3}2\Pop$.
	Assume
	that $\rho$ evolves through Eq.(\ref{eq:lindblad_generic})
	augmented by quadrature noise:
	\begin{equation}
		\frac d{dt} \rho_t = \sum_{k=0}^{2M-1} \cD[\LL_k](\rho_t)
				+ \kappa \, \cD[\qop](\rho_t)
				+ \kappa \, \cD[\Pop](\rho_t).
	\end{equation}
	Then, the expectation value of $\hop$ satisfies:
	\begin{equation}
		\frac d{dt} \trace(\hop \rho_t) = -\cA \epsilon\eta
		\trace\left( \mathcal L_\sigma h (\tfrac\eta2\qop, \tfrac\eta2\qop_2) \, \rho_t \right)
	\end{equation}
	with
	\begin{align}
		\cA =&\, e^{-\epsilon\eta/2}, \\
		\sigma \equiv& \, \sigma(\epsilon,\eta,\kappa)
		= \frac{3 \cA\epsilon\eta}8 + \frac{\kappa\eta}{8\cA\epsilon},
	\end{align}
	\begin{equation}\begin{aligned}
		\mathcal L_\sigma h =&\,
		\left( \sin(2\theta_1) + \frac12 \sin(2\theta_1+2\theta_2) - \frac12\sin(2\theta_2)\right) %
		\frac{\partial h}{\partial {\theta_1}}\\
		& + \left( \sin(2\theta_2) + \frac12 \sin(2\theta_1+2\theta_2) - \frac12\sin(2\theta_1)\right)
		\frac{\partial h}{\partial {\theta_2}}\\
		&-\sigma \left(
			\frac{\partial^2 h}{\partial \theta_1^2}
			- \frac{\partial^2 h}{\partial \theta_1 \partial\theta_2}
			+ \frac{\partial^2 h}{\partial \theta_2^2}
			\right).
		\label{eq:l_from_t_hexa}
	\end{aligned}\end{equation}
	\label{thm:eigs_hexa}
\end{thm}
The proof of Theorem~\ref{thm:eigs_hexa} relies on computations similar to
but somewhat more tedious than
the computations presented in the proof of Theorem~\ref{thm:eigs_square}
in Apprendix~\ref{adx:eigs_proof}.
Most importantly, the eigenvalues
$0 = \lambda_{0,\sigma} \leq \lambda_{1,\sigma} \leq \ldots$
of this new differential operator
$\mathcal L_\sigma$
can also be computed with the tools developed in
\cite{michelSmallEigenvalues2019a}; we find
\begin{fact}
	\begin{align}
		& \lambda_{k,\sigma} = \left(\frac{12\sqrt3}\pi +o(1)\right)\,  e^{-2/\sigma}, \quad 1\leq k\leq3,  \\
		& \exists \upepsilon_0>0, \, \forall \sigma>0,
		 \quad	\lambda_{4,\sigma} \geq \upepsilon_0.
	\end{align}
\end{fact}
\vspace{1em}

We emphasize that, even though valid only for quadrature noise,
the eigenvalue analysis led in this section serves several purposes in our study of GKP qubits.
First, having a closed-form expression of the logical decoherence rates
in a given regime provides us with insight
on the interplay between the truncation in energy and the achievable logical performance.
It is also used as a ground truth source to calibrate the parameters of
our numerical simulation scheme,
since numerical convergence can be exactly checked for the simulations with quadrature noise;
these numerical parameters can then be used for simulations with other types of noise.
On the longer term, for the simulation of multi-qubit architectures controlled with logical gates,
brute force simulations of the full Lindblad equations will prove impossible
merely due to the dimension of the Hilbert space growing exponentially with the number of qubits.
Extension of the above explicit solutions to these architectures,
even if restricted to a given noise model, could thus
prove instrumental for their analysis
(in addition to the necessary development of techniques such as model order reduction adapted to Lindblad equations,
aimed at reducing the numerical
dimensionality of the problem).

\input{figure_decoherence.tex}

%% file: figure_decoherence.tex
\begin{figure}
	\begin{center}
	\includegraphics[scale=0.60]{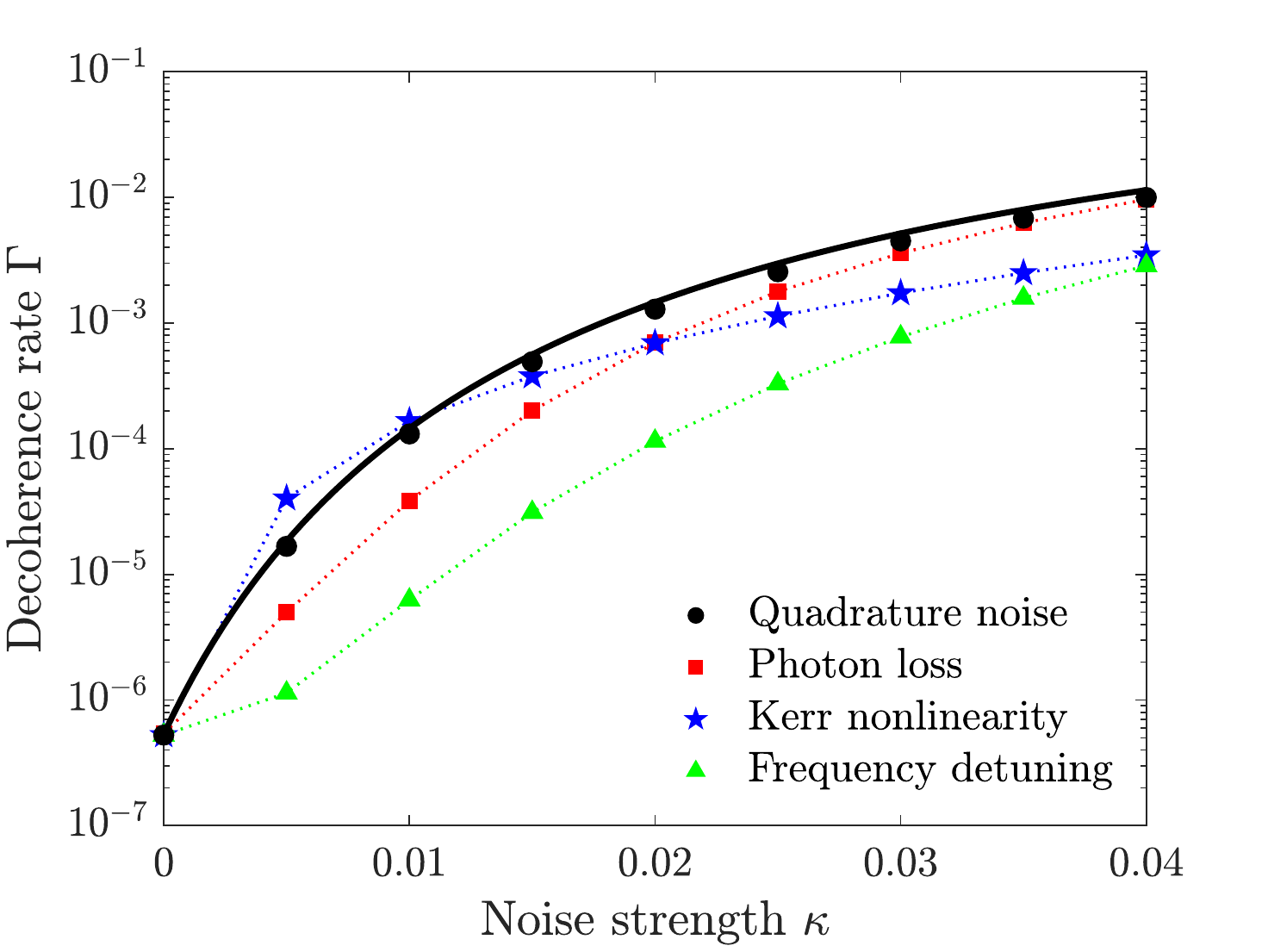}
	\caption{%
		{\bf{Logical decoherence rates associated to typical noise processes.}} %
		We simulate Eq.(\ref{eq:lindblad_with_noise}) in the square case
		($M = 2$, $\eta = 2\sqrt\pi$, and we choose $\epsilon=0.1$ here): %
		\( \frac d{dt} \rho = %
			\sum_{k=0}^3 \cD [\LL_k](\rho) %
			+ \kappa \, \mathcal L_{\textrm{noise}}(\rho) \)
		for different values of $\mathcal L_{\textrm{noise}}$, %
		modeling typical noise processes in circuit-QED experiments. %
		For each type of noise (associated to a given color),
		we compute the decoherence rate $\GGamma$ (colored dots),
		obtained by fitting the decay of the expectation value of
		$\trace(\X\rho)$ along trajectories,
		as a function of the noise strength $\kappa$.
		Colored dotted lines are only guides for the eye.
		The solid {\bf black} line corresponds to the theoretical
		asymptotic decoherence rate
		\( \GGamma_0(\kappa) = \tfrac{4\cA\epsilon\eta}\pi \, %
				e^{-\left( \frac{\cA\epsilon\eta}4 %
				+ \frac{\kappa\eta}{8\cA\epsilon} \right)^{-1} } \)
		obtained by taking $h$ as the eigenfunction of $\mathcal L_\sigma$
		associated to its first non-zero eigenvalue in
		Eq.(\ref{eq:evolution_expectation_eigs})
		and replacing the eigenvalue by the leading term
		of its asymptotic expansion in
		Eq.(\ref{eq:asymptotic_lambda1}).
		{\bf Black} dots correspond to \emph{quadrature noise}:
		\( \mathcal L_{\textrm{noise}} (\rho) = \cD[\qop](\rho) + \cD[\Pop](\rho), \)
		which is the only type of noise for which the
		method of Section~\ref{sec:explicit_rates} provides explicit decoherence rates.
		We see that these rates are well captured
		by the asymptotic rate $\GGamma_0$ in the considered range.
		{\bf \color{red} Red} squares correspond to \emph{photon loss}
		(modeling leakage in the experiment):
		\( \mathcal L_{\textrm{noise}}(\rho) = \cD[\oa](\rho) \)
		where $\oa = \tfrac{\qop + i\Pop}{\sqrt2}$ is the annihilation operator.
		{\bf \color{green} Green} triangles correspond to a \emph{detuning} Hamiltonian
		(modeling mismatched frequencies in the experiment):
		\( \mathcal L_{\textrm{noise}}(\rho) = -i \, [ \N, \rho]. \)
		{\bf \color{blue} Blue} stars correspond to a \emph{Kerr} Hamiltonian
		(modeling a type of non-linear effect):
		\( \mathcal L_{\textrm{noise}}(\rho) = -i\tfrac\epsilon\eta \, [ \N^2, \rho]. \)
		The factor $\epsilon/\eta$ is introduced to account
		for the fact that
		this process involves second-order polynomial terms in $\N$
		and is thus expected to scale differently than lower-order processes.
		We infer this scaling by a mean-field type argument,
		expecting the decoherence introduced by a Kerr Hamiltonian of strength
		$\kappa$ to be of a similar order to that introduced by a detuning Hamiltonian
		of strength $\langle \N\rangle \kappa$,
		where the average photon number
		$\langle \N\rangle$
		was shown to scale as $\eta/\epsilon$
		in Section~\ref{sec:energy_estimates}.
		A similar scaling was observed for \emph{dephasing} (not shown for clarity):
		\( \mathcal L_{\textrm{noise}}(\rho) = \tfrac\epsilon\eta \,\cD[\N](\rho) \),
		for which the same mean-field argument applies
		(recall that a dissipator is a quadratic expression of the corresponding
		Lindblad operator).
		}
	\label{fig:numerics}
	\end{center}
\end{figure}

%% file: conclusion.tex
The exponentially small decoherence rates obtained here
indicate that finite-energy GKP grid states,
stabilized by an engineered dissipative environment,
could be used to encode logical qubits
with residual logical error rates compatible with
the stringent requirements of quantum error correction.
Our numerical simulations involving different kinds of physically relevant noise processes
indicate that this protection should be robust to small experimental imperfections.

A key virtue of the model considered here is that the Lindblad dynamics to engineer
should be accessible with current circuit-QED tools,
opening the way to experimental implementation and verification of the predicted logical protection
properties.
From an experimental viewpoint, the explicit formulas for the decoherence rates in presence of
quadrature noise can serve, in first approximation,
to estimate the experimental parameter regime necessary to reach a given level
of logical protection, \emph{e.g.} in relation to the threshold of a given error correcting code
to be concatenated with the GKP encoding;
in turn, these parameter estimates can serve to identify fabrication challenges to be addressed.

Finally, in future publications,
we hope to leverage the formal \emph{a priori} energy estimates
to develop a fully rigorous analysis of the system on the one hand,
and extend our results to multi-qubit architectures on the other.

%% file: appendix_energy.tex
Using the definition of the operators $\LL_k$,
we compute the evolution of $\N$ as
\begin{align}
	\label{eq__evol_n_heisenberg}
	\frac d{dt}\trace(\N\rho_t) &= %
	\sum_{k=0}^{2M-1} \Gamma \trace(\cD^*[\LL_k](\N)\rho_t)
\end{align}
where
\begin{align}
	\cD^*[\LL](\N) &= \LL^\dag \N \LL - \frac12\left(
	\LL^\dag \LL \N + \N \LL^\dag \LL\right) \\
	&= \frac12 \, \LL^\dag \left[ \N, \LL\right] +\hc
\end{align}
and $\hc$ stands for \emph{hermitian conjugate} throughout this appendix.
As
$\N = \tfrac12\left( \qop^2 + \Pop^2 - \II\right)$
we need only compute the
evolution of $\qop^2$ and $\Pop^2$.
Using $[\qop,\Pop] = i$ we get
\begin{equation}
	\begin{split}
		[\qop^2,\LL_0]
		&= \left[ \qop^2, \cA e^{i\eta\qop} \left(\II-\epsilon\Pop\right) -\II\right] 
		= -2i\epsilon\cA \, \qop e^{i\eta\qop}
	\end{split}
\end{equation}
hence
\begin{equation}
	\begin{split}
		\LL_0^\dag [\qop^2,\LL_0]
		&= \left( \cA \left(\II-\epsilon\Pop\right) e^{-i\eta\qop} - \II\right)
		[\qop^2,\LL_0]\\
		&= -2i\epsilon\cA^2 \qop + 2i\epsilon^2 \cA^2 \Pop\qop
		+ 2i\epsilon\cA \qop e^{i\eta\qop}
	\end{split}
\end{equation}
and
\begin{equation}
	\begin{split}
		\frac12 \, \LL_0^\dag [\qop^2,\LL_0] +\hc
		&= \epsilon^2 \cA^2 \, \II - 2\epsilon\cA \, \qop\sin(\eta\qop).
	\end{split}
\end{equation}
Similarly, using also $[f(\qop),\Pop] = i \, f'(\qop)$ (for f analytic):
\begin{equation}
	\begin{split}
		[\Pop^2,\LL_0]
		&= \left[ \Pop^2, \cA e^{i\eta\qop} \left(\II-\epsilon\Pop\right) - \II\right] \\
		&= \eta\cA \,  e^{i\eta\qop}\left( 2\Pop + \eta\II\right)  \, (\II-\epsilon\Pop)
	\end{split}
\end{equation}
hence
\begin{equation}\begin{split}
	\LL_0^\dag [\Pop^2, \LL_0]
	=& \left( \cA \left(\II-\epsilon\Pop\right) e^{-i\eta\qop} - \II\right)
	[\Pop^2, \LL_0]\\
	=& \, \eta^2\cA^2 \II + 2\eta (1-\epsilon\eta) \cA^2\Pop
	+ 2\epsilon^2 \eta\cA^2 \Pop^3\\
	& -\epsilon\eta (4-\epsilon\eta)\cA^2\Pop^2
	- \eta^2\cA e^{i\eta\qop}\\
	& -\eta (2-\epsilon\eta) \cA e^{i\eta\qop}\Pop
	+ 2\epsilon\eta\cA e^{i\eta\qop}\Pop^2
\end{split}\end{equation}
and
\begin{equation}\begin{split}
	\frac12 \, \LL_0^\dag  [ & \Pop^2,\LL_0] +\hc \\
	=&\, \eta^2\cA^2 \II + 2\eta (1-\epsilon\eta) \cA^2\Pop
	-\epsilon\eta (4-\epsilon\eta)\cA^2\Pop^2 \\
	&+ 2\epsilon^2 \eta\cA^2 \Pop^3
	+\frac{\epsilon\eta^3}{2}\cA \cos(\eta\qop)
	+ 2\epsilon\eta\cA \, \Pop \cos(\eta\qop) \Pop\\
	& -\eta (2+\epsilon\eta)\cA \cos(\eta\qop)\Pop
	-i \frac{\eta^2}2 (2+\epsilon\eta)\cA \sin(\eta\qop).
\end{split}\end{equation}
Let us now define
\(
	\theta_k = \frac{k\pi}M
\)
for $k\in\mathbb Z$
and
\begin{equation}\begin{split}
	\qop_k = e^{i\theta_k\N} \, \qop \, e^{-i\theta_k\N}
	= \cos(\theta_k) \qop + \sin(\theta_k)\Pop, \\
	\Pop_k = e^{i\theta_k\N}\, \Pop \, e^{-i\theta_k\N}
	= \cos(\theta_k) \Pop - \sin(\theta_k)\qop.
\end{split}\end{equation}

With these notations, we have
\begin{equation}\begin{split}
	\cD^*[\LL_k](\N) &=
	e^{\frac{ik\pi}{M}\N} \,
	\cD^*[\LL_0](\N) \, %
	e^{-\frac{ik\pi}{M}\N}\\
	&= \frac12 \,
	e^{i\theta_k\N} \,
	\cD^*[\LL_0](\qop^2+\Pop^2) \, %
	e^{-i\theta_k\N} \\
	&= \frac12 \,
	e^{i\theta_k\N} \left( %
	\frac12 \LL_0^\dag [\qop^2+\Pop^2, \LL_0] + \hc \right)e^{-i\theta_k\N}.
\end{split}\end{equation}
Since $\qop_{k+M}=-\qop_k$, $\Pop_{k+M}=-\Pop_k$ we get
\begin{align}
	\sum_{k=0}^{2M-1} \Pop_k
	&= \sum_{k=0}^{2M-1} \Pop_k^3 =0
\end{align}
and
\begin{align}
	\sum_{k=0}^{2M-1} \cos(\eta\qop_k)\Pop_k
	&= \sum_{k=0}^{2M-1} \sin(\eta\qop_k)
	=0
\end{align}
so that
\begin{equation}\begin{split}
	\frac12	\sum_{k=0}^{2M-1} & \,
	e^{i\theta_k\N} \left( %
	\frac12 \LL_0^\dag [\qop^2+\Pop^2, \LL_0] + \hc%
	\right)e^{-i\theta_k\N}\\
	=& \, M (\epsilon^2+\eta^2) \cA^2   \, \II
	- \epsilon\cA \sum_{k=0}^{2M-1} \qop_k \sin(\eta\qop_k) \\
	& -\epsilon\eta(2-\epsilon\eta/2)\cA^2 \sum_{k=0}^{2M-1} \Pop_k^2
	+\frac{\epsilon\eta^3}{4}\cA \sum_{k=0}^{2M-1} \cos(\eta\qop_k)\\
	&+\epsilon\eta\cA \sum_{k=0}^{2M-1} \Pop_k \cos(\eta\qop_k)\Pop_k.
\end{split}\end{equation}
A straightforward computation shows that
\begin{equation}
	\sum_{k=0}^{2M-1} \Pop_k^2
	= \sum_{k=0}^{2M-1} \qop_k^2
	= M \left( \qop^2 + \Pop^2 \right).
\end{equation}
Moreover, for any $\alpha,\beta >0$ and $r\in (0,1)$
the following operator inequalities hold:
\begin{align}
	- \qop_k \sin(\eta\qop_k) & \leq |\qop_k| \\
	\Pop_k \cos(\eta\qop_k) \Pop_k & \leq \Pop_k^2 \\
	\cos(\eta\qop_k) & \leq \II \\
	-\alpha \qop_k^2 + \beta |\qop_k|
	& \leq - r\alpha \qop_k^2 + \frac{\beta^2}{4\alpha (1-r)} \II.
\end{align}
Using these inequalities
and defining
\begin{equation}
	\phi_{\epsilon,\eta} = (2-\epsilon\eta/2)\cA -1
\end{equation}
we find
\begin{equation}\begin{split}
	\frac 12 \sum_{k=0}^{2M-1}  \,
	e&^{i\theta_k\N}  \left( %
	\frac12 \LL_0^\dag [\qop^2+\Pop^2, \LL_0] + \hc%
	\right)e^{-i\theta_k\N}\\
	\leq & \, M \left( (\epsilon^2+\eta^2) \cA^2  + \frac{\epsilon\eta^3}2 \cA\right) \, \II
	+ \epsilon\cA \sum_{k=0}^{2M-1} \vert\qop_k\vert
	-\epsilon\eta\cA \, \phi_{\epsilon,\eta}
	\sum_{k=0}^{2M-1} \qop_k^2 \\
	\leq & \, M\cA \left( (\epsilon^2+\eta^2) \cA  + \frac{\epsilon\eta^3}2
	+ \frac{\epsilon}{2\eta (1-r) \phi_{\epsilon,\eta}}
	\right) \, \II
	- r\, \epsilon\eta\cA \phi_{\epsilon,\eta}%
	\sum_{k=0}^{2M-1} \qop_k^2 \\
	=&\, M\cA \Big( (\epsilon^2+\eta^2) \cA  + \frac{\epsilon\eta^3}2
	+ \frac{\epsilon}{2\eta (1-r) \phi_{\epsilon,\eta}}
	- r\, \epsilon\eta \phi_{\epsilon,\eta}
	\Big) \, \II 
	- 2 M \cA \, r\, \epsilon\eta \phi_{\epsilon,\eta} \, \N.
\end{split}
\end{equation}

Hence the operator inequality
\begin{equation}\begin{split}
	\Gamma \sum_{k=0}^{2M-1} %
	\cD^*[\LL_k](\N)  %
	\leq -\lambda(r,\epsilon,\eta)\N + \mu(r,\epsilon,\eta)\II
\end{split}\end{equation}
with
$\lambda$ and $\mu$ defined in Eq.(\ref{eq__lambda_coef})--(\ref{eq__mu_coef})
and corresponding respectively to the prefactor before $\N$ and $\II$
in the final inequality above.

%% file: appendix_eigs.tex
We start by studying the noiseless case
and explain at the end of this appendix
how to take quadrature noise into account.
The Lindblad dynamics is thus given by:
\begin{equation}
	\frac d{dt}\rho = \Gamma \, \sum_{k=0}^3 \cD[\LL_k](\rho)
\end{equation}
where
$\epsilon>0$,
$\cA = e^{-\epsilon\eta/2}$,
\(
	\cD[\LL](\rho) = \LL\rho\LL^\dag - \frac12\left( \LL^\dag\LL \rho + \rho\LL^\dag\LL\right)
\)
and
\begin{equation}
	\LL_k = \cA
		e^{ik\tfrac{\pi}{2}\N}
		\, e^{i\eta\qop} \left( \II-\epsilon\Pop\right) \,
		e^{-ik\tfrac{\pi}{2}\N}
		- \II.
\label{eq:eigs_lindblad_square}
\end{equation}

Let us now compute the evolution under Eq.(\ref{eq:eigs_lindblad_square})
of any
separable periodic observable of the form
\begin{equation}
	\hop = h(\qop,\Pop) = f(\tfrac\eta2\qop) \, g(\tfrac\eta2\Pop)
\end{equation}
where $f$ and $g$ are smooth real-valued $2\pi$-periodic functions:
\begin{equation}
	\begin{aligned}
		\frac d{dt} \trace(\hop\rho_t)
		&= \Gamma \sum_{k=0}^3 \trace\left( \cD^*[\LL_k](\hop) \, \rho_t \right)
	\end{aligned}
\end{equation}
where
\begin{align}
	\cD^*[\LL_k](\hop) &=
		\LL_k^\dag \, \hop \, \LL_k
		-\frac12\left( \LL_k^\dag\LL_k \, \hop + \hop \, \LL_k^\dag \LL_k\right)\\
		&= \frac12 \left( \LL_k^\dag \, [\hop, \LL_k] + [\LL_k^\dag , \hop] \, \LL_k \right).
\end{align}
Crucially, expanding $f$ and $g$ in Fourier series and using the Baker-Campbell-Hausdorff formula,
we see that
\begin{align}
[f(\tfrac\eta2\qop), e^{i\eta\Pop}] = [g(\tfrac\eta2\Pop), e^{i\eta\qop}] = 0.
\label{eq:commutation_f_g_exp}
\end{align}
Using Eq.(\ref{eq:commutation_f_g_exp})
and the formula $[f(\qop), \Pop] = if'(\qop)$ (for $f$ analytic)
we get:
	\begin{align}
		\LL_0^\dag \, [\hop, \LL_0]
		&= \LL_0^\dag \, [f(\tfrac\eta2\qop)\, g(\tfrac\eta2\Pop),
			\mathcal A \, e^{i\eta\qop} \, \left( \II - \epsilon \Pop\right) - \II]\notag\\
		&= - \mathcal A\epsilon \, \LL_0^\dag \, e^{i\eta\qop} \,
		[f(\tfrac\eta2\qop)\, ,\Pop] \, g(\tfrac\eta2\Pop)\notag\\
		&= -i\, \frac{\mathcal A\epsilon\eta}2 \, \LL_0^\dag \, e^{i\eta\qop} \,
		f'(\tfrac\eta2\qop) \, g(\tfrac\eta2\Pop)\notag\\
		&= -i\, \frac{\mathcal A\epsilon\eta}2 \,
		\left( \mathcal A (\II-\epsilon\Pop) - e^{i\eta\qop}\right) \,
		f'(\tfrac\eta2\qop) \, g(\tfrac\eta2\Pop),\\
		 \notag\\
		\relax [\LL_0^\dag , \hop]\, \LL_0
		&= i \, \frac{\mathcal A \epsilon\eta}2\, f'(\tfrac\eta2\qop)\, g(\tfrac\eta2\Pop)\,
			\left( \mathcal A (\II-\epsilon \Pop) - e^{-i\eta\qop}\right),\\
		 \notag\\
		\cD^*[\LL_0] (\hop)
		&=\, \frac12 \left( \LL_0^\dag \, [\hop, \LL_0] + [\LL_0^\dag , \hop]\, \LL_0 \right) \notag\\
		&=\, -i\, \frac{\mathcal A^2 \epsilon^2\eta}4\,
			[ f'(\tfrac\eta2\qop) \, g(\tfrac\eta2\Pop), \Pop]
		   + i\, \frac{\mathcal A \epsilon\eta}2 \, f'(\tfrac\eta2\qop)\, g(\tfrac\eta2\Pop) \,
		   \frac{e^{i\eta\qop} - e^{-i\eta\qop}}2\notag\\
		&=\, \left(
			\frac{\mathcal A^2 \epsilon^2\eta^2}8\,
			f''(\tfrac\eta2\qop)
		   - \frac{\mathcal A \epsilon\eta}2 \, \sin(\eta\qop) f'(\tfrac\eta2\qop)
		   \right) g(\tfrac\eta2\Pop).
	\end{align}
Similar computations lead to
\begin{equation}
	\begin{aligned}
		\cD^*[\LL_2]&(\hop) = \cD^*[\LL_0](\hop),\\
		\cD^*[\LL_1]&(\hop) = \cD^*[\LL_3](\hop) \\
		&= f(\tfrac\eta2\qop) \, \left(
			\frac{\mathcal A^2 \epsilon^2\eta^2}8\,
			g''(\tfrac\eta2\Pop)
		   - \frac{\mathcal A \epsilon\eta}2 \, \sin(\eta\Pop) g'(\tfrac\eta2\Pop) \right)
	\end{aligned}
\end{equation}
so that
\begin{equation}\label{eq:lindblad_heisenberg}\begin{aligned}
	\Gamma \sum_{k=0}^3 \cD^* [\LL_k] (\hop) =
	& -\mathcal A\epsilon\eta \Gamma \,
		\left(
		\sin(\eta\qop) f'(\tfrac\eta2\qop)
		- \frac{\mathcal A\epsilon\eta}4 f''(\tfrac\eta2\qop) \right) g(\tfrac\eta2\Pop) \\
	& -\mathcal A\epsilon\eta \Gamma \,
		f(\tfrac\eta2\qop)\left(
		\sin(\eta\Pop) g'(\tfrac\eta2\Pop)
		- \frac{\mathcal A\epsilon\eta}4 g''(\tfrac\eta2\Pop) \right)
	\\
	&= -\mathcal A\epsilon\eta \Gamma \,
	\mathcal L_\sigma (h)(\tfrac\eta2\qop, \tfrac\eta2\Pop)
\end{aligned}\end{equation}
where $\sigma = \frac{\mathcal A\epsilon\eta}4$
and $\mathcal L_\sigma$ is the differential operator defined by
\begin{equation}\begin{aligned}
	\label{eq:lsigmatensor}
	\mathcal L_\sigma &= \mathcal T_\sigma\otimes \II + \II\otimes \mathcal T_\sigma,\\
	(\mathcal T_\sigma f) (\theta) &= \sin(2\theta) f'(\theta) - \sigma f''(\theta).
\end{aligned}\end{equation}

Remarkably,
the previous exact eigenvalue analysis can be adapted to take into account the effect of
quadrature noise,
\emph{i.e.}
additional dissipators in $\qop$ and $\Pop$
in the studied Lindblad dynamics.
Indeed, take a separable periodic observable
\(
	\hop = h(\tfrac\eta2\qop,\tfrac\eta2 \Pop)
	= f(\tfrac\eta2\qop) \, g(\tfrac\eta2 \Pop)
\)
and compute
\begin{equation}
	\begin{aligned}
		\frac d{dt} \trace\Big( \hop &
			\big( \kappa \cD[\qop](\rho_t) + \kappa \cD[\Pop](\rho_t)\big)
			\Big)\\
		&= \kappa \trace\left( \cD^*[\qop](\hop)\rho_t\right)
		+ \kappa \trace\left( \cD^*[\Pop](\hop)\rho_t\right).
	\end{aligned}
\end{equation}
We get
	\begin{align}
		\qop^\dag \, [\hop, \qop]
		&= \qop [\hop,\qop]
		= -i\frac\eta2 \, \qop \, f(\tfrac\eta2\qop) \, g'(\tfrac\eta2\Pop),\\
		[\qop^\dag , \hop] \, \qop
		&= [\qop, \hop]\,\qop
		= i\frac\eta2 \, f(\tfrac\eta2\qop) \, g'(\tfrac\eta2\Pop)\, \qop, \\
		\cD^*[\qop](\hop)
		&= \frac12 \left(
		\qop^\dag \, [\hop, \qop] + [\qop^\dag , \hop] \, \qop
		\right) \notag \\
		&= -i\frac\eta4 \left[ \qop, f(\tfrac\eta2\qop)\, g'(\tfrac\eta2\Pop) \right] \notag \\
		&= \frac{\eta^2}8 f(\tfrac\eta2\qop) \, g''(\tfrac\eta2\Pop),\\
		\cD^*[\Pop](\hop)
		&= \frac{\eta^2}8 f''(\tfrac\eta2\qop) \, g(\tfrac\eta2\Pop).
	\end{align}
Note that these terms already appeared in Eq.(\ref{eq:lindblad_heisenberg})
so that putting everything together we obtain
\begin{equation}
	\begin{aligned}
	\Gamma \sum_{k=0}^3 \cD^*[\LL_k](\hop) &
		+ \kappa \left( \cD^*[\qop](\hop) + \cD^*[\Pop](\hop) \right)
		= -\mathcal A\epsilon\eta \Gamma \,
	\mathcal L_\sigma (h)(\tfrac\eta2\qop, \tfrac\eta2\Pop)
	\end{aligned}
\end{equation}
where the differential operator $\mathcal L_\sigma$ is still defined by Eq.(\ref{eq:lsigmatensor})
but its parameter $\sigma$ now also depends on $\kappa$
through
\begin{equation}
	\sigma = \frac{\mathcal A\epsilon\eta}4 + \frac{\kappa\eta}{8\mathcal A\epsilon\Gamma}.
\end{equation}